\documentclass[aps,prl,reprint,superscriptaddress,
               amsmath,amssymb,amsfonts]{revtex4-2}

\usepackage{overpic}
\usepackage[varg]{txfonts}
\usepackage{microtype}
\usepackage{bm}
\usepackage[nolist]{acronym}
\usepackage{braket}
\usepackage{graphicx}
\usepackage{xcolor}

\definecolor{darkerblue}{RGB}{0,0,192}
\usepackage[colorlinks,
            linkcolor=darkerblue,    
            citecolor=darkerblue,  
            urlcolor=darkerblue]{hyperref}

\renewcommand{\vec}[1]{\boldsymbol{#1}}
\newcommand{\vhat}[1]{\hat{\vec{{#1}}}}
\newcommand{\mat}[1]{\boldsymbol{#1}}
\newcommand{\Ap}{\mathcal{A}}

\newcommand{\beq}{\begin{equation}}
\newcommand{\eeq}{\end{equation}}

\newcommand{\avg}[1]{\braket{#1}}
\newcommand{\trp}[1]{{#1}^{\intercal}}
\newcommand{\cc}[1]{{#1}^*}
\newcommand{\h}[1]{{#1}^{\dagger}}
\newcommand{\nh}[1]{{#1}^{\phantom \dagger}}

\newenvironment{subalign}{\subequations\align}{\endalign\endsubequations}

\newcommand{\Pin}{\vec{P}_{\rm in}}
\newcommand{\Pout}{\vec{P}_{\rm out}}

\newcommand{\hatPin}{\vhat{P}_{\rm in}}

\newcommand{\Jo}{J}

\newcommand{\Jt}{J_{110}}
\newcommand{\Jtp}{J_{110}'}

\newcommand{\Jf}{J_{121}}
\newcommand{\Jfp}{J_{121}'}

\newacro{SOC}[SOC]{spin-orbit coupling}    

\begin{document}

\title{Observing Altermagnetism using Polarized Neutrons}

\author{Paul~A.~McClarty}
\affiliation{Laboratoire L\'{e}on Brillouin, CEA, CNRS,  Universit\'{e} Paris-Saclay, CEA-Saclay, 91191 Gif-sur-Yvette, France}
\email{paul.mcclarty@cea.fr}
\author{Arsen Gukasov}
\affiliation{Laboratoire L\'{e}on Brillouin, CEA, CNRS,  Universit\'{e} Paris-Saclay, CEA-Saclay, 91191 Gif-sur-Yvette, France}
\email{arsen.goukassov@cea.fr}
\author{Jeffrey~G.~Rau}
\affiliation{Department of Physics, University of Windsor, 401 Sunset Avenue, Windsor, Ontario, N9B 3P4, Canada}
\email{jrau@uwindsor.ca}


\begin{abstract}
Altermagnets are colinear compensated magnets whose magnetic symmetries at zero spin-orbit coupling break spin degeneracy leading to spin-split electronic and magnonic bands that reflect an underlying multipolar order. When there is an approximate $U(1)$ symmetry the magnons in altermagnets are split into equal and opposite chiral pairs. 
We show that in altermagnets polarized neutrons
provide a means to detect the population of time-reversed domains and allow direct measurement of the magnon chirality anisotropy in momentum space -- the central signature of the altermagnetic phase. We demonstrate this response to polarized neutrons in two candidate materials MnF\textsubscript{2} and MnTe and show that the presence of these chiralities is stable to small perturbations that break spin-rotation symmetry. This provides a magnonic analogue of spin polarized ARPES that has been used to discern altermagnetism in the electronic band structures of various candidate materials. 
\end{abstract}

\maketitle

Many of the advances in condensed matter physics of this century have originated from a focus on systems with strong spin-orbit coupling. This has been true of much of the broader field of band topology as well as in research into exotic superconductivity and the collective phenomena of magnetic systems. In recent years there has been renewed interest in systems where the spin-orbit coupling is \emph{weak} driven, in part, by the presence of generalized symmetries such as spin-space symmetries \cite{corticelli2022,liu,schiff2023,yang2023symmetry,ren2023enumeration,xiao2023spin,jiang2023enumeration}. This is exemplified by the discovery of altermagnetism~\cite{Smejkal2019,hayami2019,Smejkal2022b} where the conditions for spin degeneracy are broken in antiferromagnets at zero spin-orbit coupling as a direct consequence of their distinctive magneto-crystalline symmetries. 

Altermagnets, as defined from the zero spin-orbit coupled limit, have the feature that their electronic bands are spin split and feature an anisotropy in momentum space \cite{Smejkal2019,hayami2019,Smejkal2022b} reflecting an underlying multipolar order~\cite{Mcclarty2024,bhowal2024}. Recently, ARPES and its spin-polarized analog have been used to directly visualize the breaking of spin degeneracy along momentum space cuts in altermagnetic candidate materials such as MnTe \cite{lee2024,krempaski2024}. 

Signatures of altermagnetism are also visible via the magnetic excitations~\cite{Smejkal2023,Gohlke2023}. In a classic Heisenberg-N\'{e}el antiferromagnet with two magnetic sublattices there is a single doubly degenerate magnon band due to a good quantum number defining a  magnon ``chirality''~\cite{lovesey}. The existence of this conserved chirality is little more than a curiosity under these circumstances, with weak spin-orbit coupling often favoring states with zero net chirality. While in an antiferromagnet these chiralities can only be split using (e.g.) a magnetic field, in an altermagnet the magnon chiralities are intrinsically split in energy due to an isotropic exchange over most of the Brillouin zone -- exhibiting the \emph{same} anisotropy in momentum space as the one present in their electronic bands. Experimental evidence for this altermagnetic splitting has been observed in MnTe~\cite{liu2024} and should be observable in MnF\textsubscript{2} through measurements along specific high-symmetry directions~\cite{corticelli2022}. 
\begin{figure*}
    \centering    
    \begin{overpic}[width=0.8\columnwidth]{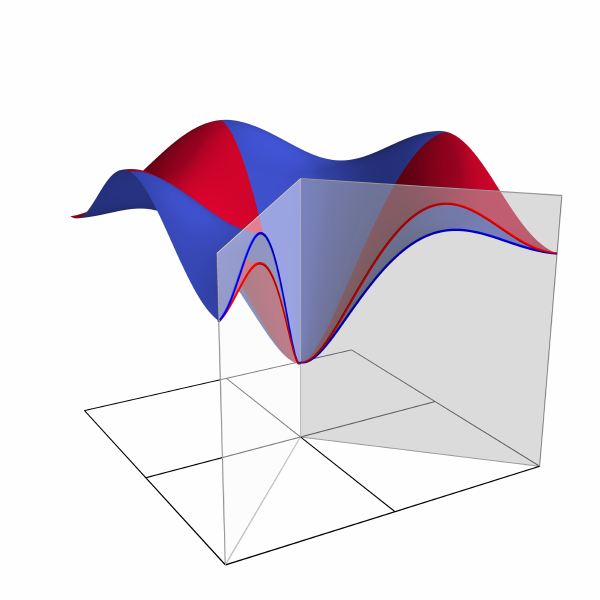}
    \put(10,90){(a)}
    \put(35,2){$[1\bar{1}0]$}
    \put(92,20){$[1{1}0]$}
    \put(66,11){$k_x$}
    \put(75,33){$k_y$}
    \put(15,80){\textcolor{red}{$P^z_{\rm out}>0$}}
    \put(15,50){\textcolor{blue}{$P^z_{\rm out}<0$}}
    \put(50,77){$E/(\Jo S)$}
    \end{overpic}
    \hspace{2em}
    \includegraphics[width=\columnwidth]{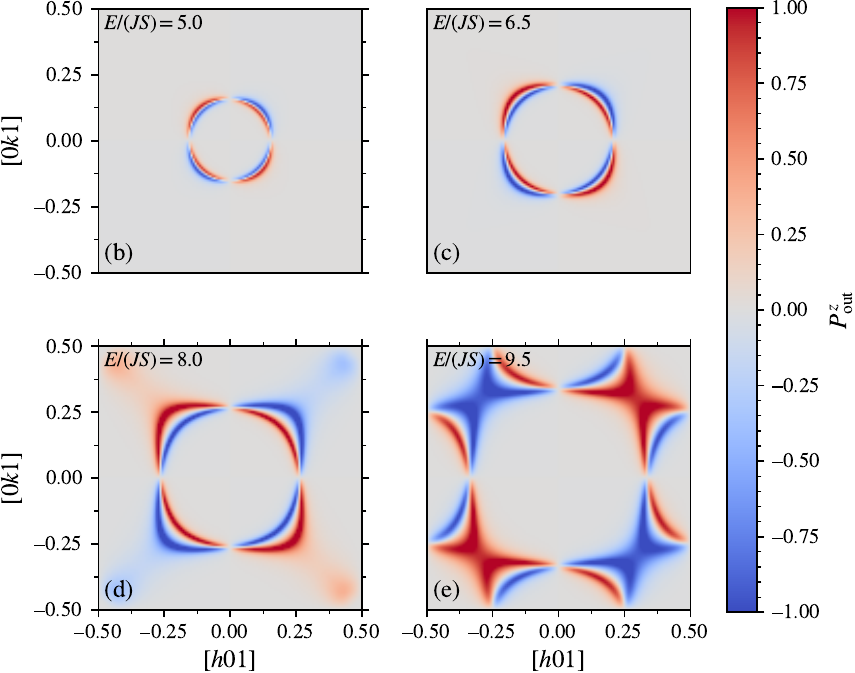}    
    \caption{(a) Illustration of the magnon dispersion of MnF\textsubscript{2} ($\Jt/\Jo = -0.25$, $\Jtp/\Jo = -0.5$) with finite altermagnetic exchange differences in the $[h,k,0.25]$ plane. The order parameter is oriented along $\vhat{z}$. The chirality of the magnon bands, which determines the polarization of outgoing neutrons $P^z_{\rm out}$, is indicated by the color of the bands. (b-e) Cuts in the $[hk1]$ plane at fixed energy $E/(\Jo S)$ of the dispersion of the MnF\textsubscript{2} magnon bands showing the polarization of the outgoing neutron at each energy and wavevector. Polarization is averaged with a (Lorentzian) energy broadening $\delta E/\Jo S) = 0.25$ to loosely mimic finite experimental resolution.
    }
    \label{fig:mnf2}
\end{figure*}
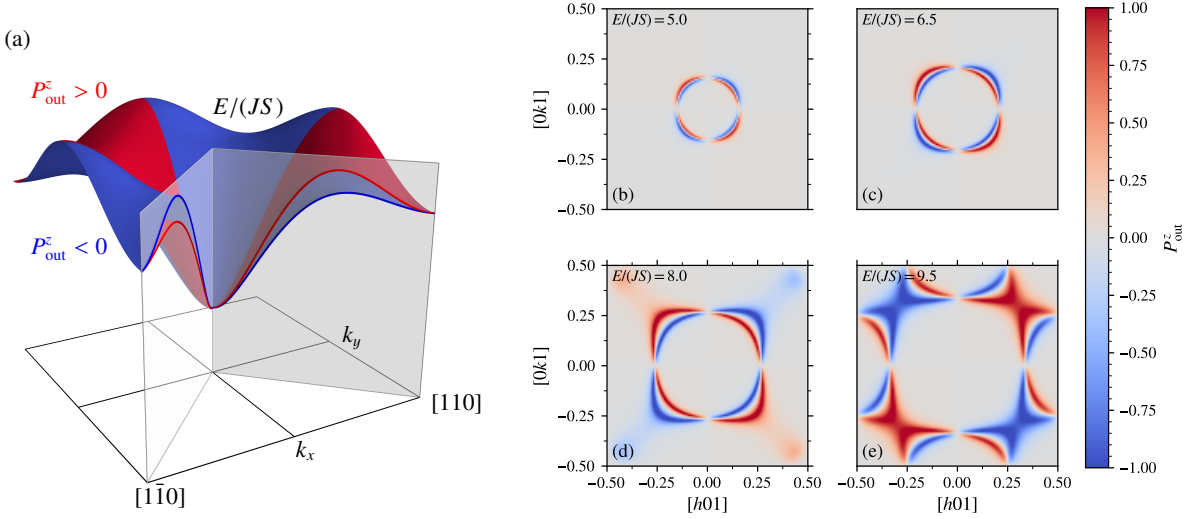

In this paper, we point out that neutron scattering provides the means to distinguish all altermagnetic magnetic domains, including those related by time reversal, and the chiralities of the magnon modes in altermagnets. We will show that this is a \emph{general} feature of altermagnets -- a direct consequence of their defining symmetries -- and that it is stable to the inclusion of weak spin-orbit coupling.
The key to these measurements is the ability to control the polarization of the ingoing neutron beam and to measure the polarization of the scattered beam. 
It is often said that altermagnets blend certain properties of ferromagnets and antiferromagnets. As we show, a majority single-domain altermagnet polarizes an initially unpolarized neutron beam through inelastic scattering from its magnon excitations. 
In this sense altermagnets behave like ferromagnets~\cite{schwinger1937} despite being entirely compensated like antiferromagnets. 
However, the direction of polarization has a much richer dependence  on the local moment orientation, magnon band and scattering wavevector than in simple ferromagnets. 
Experiments using polarized neutrons can thus be used to map out the anisotropic splitting of chirality, just as spin-polarized ARPES can be used to map out the anisotropic pattern of spin splitting of the electronic band structures of altermagnets. 
Going further, the relative intensities of each of the chiral modes can be modulated across the Brillouin zone by tuning the polarization of the ingoing neutron beam.
By taking data with flipped ingoing polarizations it is possible to isolate and measure the chiralities of the magnon bands individually, mirroring the signature in the outgoing polarization.

Observability of these effects rests on having a sample with an imbalance in the populations of domains related by time reversal -- ideally a single magnetic domain. In some systems this can be practically achieved by annealing from a high-field state (as has been done in MnF\textsubscript{2} ~\cite{Alperin1962,Felcher1996}).
We give a general argument that, in altermagnets, (elastic) neutron polarimetry can distinguish such domains thus providing the means to identify or validate suitable samples for inelastic scattering experiments using polarized neutrons. 
Whereas previous work has demonstrated experimentally that magnon chirality can be probed using polarized neutrons in uncompensated magnets~\cite{nambu2020,jenni2022}, this work establishes neutron polarization as a powerful tool to characterize altermagnets generally through their magnetic excitations. 

We organize the rest of the paper as follows. 
We first explain how the ratio of time reversed domains in altermagnets can be measured by exploiting nuclear-magnetic interference in neutron diffraction. 
We then give a brief introduction to altermagnetic splittings of chiral magnon modes including examples of toy models for candidate altermagnetic materials MnF\textsubscript{2} and MnTe. 
Turning then to inelastic neutron scattering of polarized neutrons, we show that the chiralities of altermagnetic magnons can be measured in the spin-orbit free limit. 
Finally, we discuss the role of spin-orbit coupling as a perturbation to this limit which is potentially an important consideration when carrying out studies of real altermagnetic materials. 

{\it Diffraction } $-$ The cross section for neutron diffraction using unpolarized neutrons is time reversal invariant and therefore they cannot be used to distinguish magnetic domains related by time reversal symmetry. However, when the neutrons are polarized, there is a contribution in the cross section that is time reversal odd originating from the interference of nuclear and magnetic contributions. This contribution can, in principle, distinguish such magnetic domains. 

This point was made already in 1962~\cite{Alperin1962} for the case of MnF$_2$. The observation in that work rests on the identification of the crystalline symmetries now central to defining altermagnetism. The generalization of their argument to all colinear centrosymmetric altermagnets is as follows. 

\newcommand{\axis}{\vhat{N}}
For this discussion it will be sufficient to consider moments aligned or anti-aligned along a common axis $\axis$. Then the relevant contribution to the elastic cross section at a reciprocal lattice vector, $\vec{G}$, of the crystal lattice is proportional to
\begin{equation}
\propto 
\left(\axis \cdot \Pin^\perp\right) {\rm Re} \left[ F^{\phantom *}_{\rm nuc}(\vec{G}) F^*_{\rm mag}(\vec{G}) \right].
\label{eq:NMmix}
\end{equation}
Here $\Pin^\perp \equiv \hat{\vec{G}} \times (\Pin \times \hat{\vec{G}})$ with $\Pin$ the polarization of the ingoing neutron beam. The contribution from the basis of sites in the primitive cell is $F_{\rm nuc}(\vec{G}) = \sum_{n} \exp(i\vec{G}\cdot \boldsymbol{\delta}_n) b_{n}$ where $b_{n}$ is the scattering length of the $n$\textsuperscript{th} nucleus in the unit cell and $\vec{\delta}_n$ is its position. The magnetic contribution is similar with
$
F_{\rm mag}(\vec{G}) = 
\sum_{n} \mu_n \sigma_n f_n(\vec{G})  \exp(i\vec{G}\cdot \vec{\delta}_n)
$ where $\sigma_n = \pm 1$ specifies the moment direction of each magnetic atom with respect to $\axis$ and its $\mu_n$ size, while $f_n(\vec{G})$ is its magnetic form factor.

An altermagnet can be thought of as a colinear compensated magnet where the oppositely oriented magnetic sublattices \emph{cannot} be connected by time reversal combined with either translation or inversion. Instead they are connected, for example, by the combined operations of translation, \emph{rotation}, and time reversal. Consider first the conventional case, where the ions in the primitive cell can be grouped into two sets related by a translation $\boldsymbol{\tau}$ and time reversal. Then the magnetic part, containing only the magnetic ions, takes the form 
$
F^*_{\rm mag}(\vec{G}) = 
\Phi_{\rm mag}(\vec{G}) [1-\exp(-i\vec{G}\cdot\boldsymbol{\tau})]
$ where $\Phi_{\rm mag}(\vec{G})$ is real. The nuclear contribution, takes a similar form with 
$
F_{\rm nuc}(\vec{G}) = 
\Phi_{\rm nuc}(\vec{G}) [1+\exp(i\vec{G}\cdot\boldsymbol{\tau})]
$. 
The product has vanishing real part so the domains related by time reversal cannot be distinguished using polarized neutrons. 
Instead consider the case where are two groups related by time reversal and inversion. Now  $F^*_{\rm mag}(\vec{G}) = \cc{\Phi}_{\rm mag}(\vec{G}) -  \cc{\Phi}_{\rm mag}(-\vec{G})$ and $F_{\rm nuc}(\vec{G}) = \Phi_{\rm nuc}(\vec{G}) + \Phi_{\rm nuc}(-\vec{G}) $. As $\Phi_{\rm nuc}(\vec{G})$ and $\Phi_{\rm mag}(\vec{G})$ are sums of phases perhaps weighted with (real-valued) scattering lengths or moment lengths (etc), for both $\Phi(\vec{G}) = \cc{\Phi}(-\vec{G})$ and so, again, the real part vanishes.  

To see that this part of the cross section can indeed be non-vanishing in an altermagnet we take the elementary example of MnF$_2$. This is tetragonal with primitive lattice vectors $\vec{a}_1 = a\vhat{x}$, $\vec{a}_2=a \vhat{y}$ and $\vec{a}_{3}=c\vhat{z}$. Using these basis vectors the magnetic ions lie at $(0,0,0)$ and $(1/2,1/2,1/2)$ while the fluoride ions lie at $(\pm \epsilon ,\pm \epsilon,0)$ and  $(1/2\pm \epsilon ,1/2\mp \epsilon,1/2)$. The nuclear-magnetic contribution to the scattering intensity is then from Eq.~\ref{eq:NMmix}, 
$$
{\rm Re} \left[ F_{\rm nuc}(\vec{G}) F^*_{\rm mag}(\vec{G}) \right] \propto \mu \sin(a G_x \epsilon) \sin(a G_y \epsilon) (1-\cos(\vec{G}\cdot\vec{\delta})).
$$
The Bragg reflections for this lattice are indexed by $\vec{G}=2\pi(G_1\vhat{x}/a +G_2\vhat{y}/b +G_3\vhat{z}/c)$ for $(G_1,G_2,G_3)\in\mathbb{Z}$. The nuclear-magnetic scattering is then non-vanishing for $G_1+G_2+G_3$ odd. Therefore, if we measure the Bragg intensities at these reflections $I_{\pm}$ for both flipped polarizations $\pm \vec{P}^\perp_{\rm in}$ then a non-vanishing net polarization (asymmetry) $(I_+ - I_-)/(I_+ + I_-)$ signals the presence of one majority time-reversed domain.  

As a second example, we consider altermagnetic candidate MnTe for which the Mn atoms are at positions $\vec{\delta}_{\rm Mn,1} = (0,0,0)$ and $\vec{\delta}_{\rm Mn,2}=(0,0,1/2)$ (with equal and opposite moments) and the Te atoms have positions $\vec{\delta}_{\rm Te,1}=(1/3,2/3,1/4)$ and $\vec{\delta}_{\rm Te,2}=(2/3,1/3,3/4)$ in terms of the usual hexagonal basis vectors ($\vec{a}_1 = a\vhat{x}$, $\vec{a}_2 = -a/2\vhat{x} +\sqrt{3}a/2\vhat{y}$ and $\vec{a}_3 = c\vhat{z}$). Now, one finds that
$$
{\rm Re} \left[ F_{\rm nuc}(\vec{G}) F^*_{\rm mag}(\vec{G}) \right] \propto \mu \left( \cos( \vec{G}\cdot\vec{\delta}_{\rm Te,1}) + \cos(\vec{G}\cdot\vec{\delta}_{\rm Te,2})
\right),
$$
when $\vec{G}\cdot \vec{b}_3 = G_3$ is odd and zero otherwise. Thus the nuclear-magnetic polarization dependent scattering at reflections with $G_3$ odd can distinguish the direction of the altermagnetic order.

{\it Altermagnetic magnons } $-$ The peculiar symmetries of altermagnets described above in the zero spin-orbit limit have two central consequences. One is to preserve the spin quantum number of the electronic bands and the chirality of the magnon bands. The second, originating from the replacement of time reversal symmetry by time reversal in conjunction with a point group operation, is to lift the degeneracy of the spins and chiralities in an anisotropic pattern in reciprocal space. In a $d$-wave altermagnet, for example, a four-fold rotation in momentum at fixed energy reverses both spins and magnon chiralities. 

\newcommand{\sA}{{\sf A}}
\newcommand{\sB}{{\sf B}}
We can see how these features arise in a simple setting: a two-sublattice colinear compensated magnet with a $U(1)$ symmetry along the moment direction $\vhat{N}$. The Holstein-Primakoff bosons on the two sublattices (labelled $\sA$ and $\sB$) transform under this $U(1)$ symmetry as
\begin{align*}
    a_{\vec{k},\sA} &\rightarrow e^{+i\theta} a_{\vec{k},\sA}, &
    a_{\vec{k},\sB} &\rightarrow e^{-i\theta} a_{\vec{k},\sB}. 
\end{align*}
At the level of linear spin-waves, this restricts the Hamiltonian 
to have the form
\newcommand{\F}{A}
\newcommand{\G}{B}
\begin{align*}
H = 
    \sum_{\vec{k}} \left[
    {\F}^{\sA}_{\vec{k}} \h{a}_{\vec{k},\sA} \nh{a}_{\vec{k},\sA} +
    {\F}^{\sB}_{\vec{k}} \h{a}_{\vec{k},\sB} \nh{a}_{\vec{k},\sB} +
    \left(
    {\G}^{\sA\sB}_{\vec{k}} \h{a}_{\vec{k},\sA}\h{a}_{-\vec{k},\sB}+
    {\rm h.c.}  \right)
    \right].
\end{align*}
It is natural write ${\F}^\sA_{\vec{k}}$ and ${\F}^\sB_{\vec{k}}$ equivalently as ${\F}^\sA_{\vec{k}} = {\F}_{\vec{k}} + \delta {\F}_{\vec{k}}/2$ and ${\F}^\sB_{\vec{k}} = {\F}_{\vec{k}} - \delta {\F}_{\vec{k}}/2$ as well as defining ${\G}^{\sA\sB}_{\vec{k}} = {\G}^{\sB\sA}_{-\vec{k}} \equiv {\G}_{\vec{k}}$. If we further assume an inversion symmetry, as is present in many of the canonical examples of altermagnets, then ${\F}_{\vec{k}}={\F}_{-\vec{k}}$ and $\cc{\G}_{\vec{k}} = {\G}_{-\vec{k}} = {\G}_{\vec{k}}$. The relevant Bogoliubov matrix is then~\cite{lovesey,blaizot:1986}
$$
 \mat{M}_{\vec{k}} =
  \left(
    \begin{array}{cccc}
    {\F}_{\vec{k}} +\frac{1}{2}\delta {\F}_{\vec{k}} & 0 & 0 & {\G}_{\vec{k}} \\
    0 & {\F}_{\vec{k}}-\frac{1}{2}\delta {\F}_{\vec{k}} & {\G}_{\vec{k}} & 0 \\
    0 & {\G}_{\vec{k}} & {\F}_{\vec{k}} +\frac{1}{2}\delta {\F}_{\vec{k}}& 0 \\
    {\G}_{\vec{k}} & 0 & 0 & {\F}_{\vec{k}} -\frac{1}{2}\delta {\F}_{\vec{k}}
    \end{array}
    \right).
$$
Due to the $U(1)$ symmetry, we obtain two independent blocks diagonalization yields the spin-wave energies
\begin{subalign}
    \epsilon_{\vec{k},1} &= 
    +\frac{1}{2}\delta {\F}_{\vec{k}} +
    \sqrt{{\F}_{\vec{k}}^2 - {\G}_{\vec{k}}^2} \equiv \Omega_{\vec{k}} + \frac{1}{2}\delta {\F}_{\vec{k}}, \\
    \epsilon_{\vec{k},2} &= 
    -\frac{1}{2}\delta {\F}_{\vec{k}} +
    \sqrt{{\F}_{\vec{k}}^2 - {\G}_{\vec{k}}^2} \equiv
    \Omega_{\vec{k}} - \frac{1}{2}\delta {\F}_{\vec{k}}.
\end{subalign}
The splitting of the two magnon bands is thus given by $|\epsilon_{\vec{k},1}-\epsilon_{\vec{k},2}| = |\delta {\F}_{\vec{k}}|$ and thus vanishes when symmetry requires ${\F}^\sA_{\vec{k}} = {\F}^\sB_{\vec{k}}$. In an altermagnet there is no such symmetry -- the operations linking the sublattices involve a spatial operation -- and thus $\delta {\F}_{\vec{k}}$ is generally non-zero. If this symmetry operation is denoted as $R$, then this implies $\delta A_{R(\vec{k})} = -\delta A_{\vec{k}}$ -- following the same anisotropy present in the spin-splitting as the electronic bands.

Importantly, the associated eigenvectors are not affected by $\delta {\F}_{\vec{k}}$, since its contribution is proportional to the identity in each Bogoliubov sub-block. The eigenvectors associated with each positive energy mode can thus be written as
\begin{align}
    u_{\vec{k}} &= \sqrt{\frac{\Omega_{\vec{k}} + {\F}_{\vec{k}}}{2\Omega_{\vec{k}}}}, &
    v_{\vec{k}} &= -\frac{{\G}_{\vec{k}}}{\sqrt{2\Omega_{\vec{k}}(\Omega_{\vec{k}}+{\F}_{\vec{k}})}},
\end{align}
which satisfy $|u_{\vec{k}}|^2 - |v_{\vec{k}}|^2=1$.
The standard (unpolarized) neutron dynamical structure factor can be expressed in terms of these quantities. The (textbook) result is that one-magnon intensity~\cite{lovesey} is $\propto [1 + (\vhat{N}\cdot\vhat{k})^2 ] C_{\vec{k}}$ where
\begin{equation}
\label{eq:defn-c}
C_{\vec{k}} = \sqrt{ \frac{ {\F}_{\vec{k}}+{\G}_{\vec{k}} }{{\F}_{\vec{k}}-{\G}_{\vec{k}} } }.
\end{equation}
The intensity of these two chiral modes are thus identical when ignoring neutron polarization, even when a non-zero altermagnetic splitting is present.

\begin{figure}
\centering
    \includegraphics[width=0.9\columnwidth]{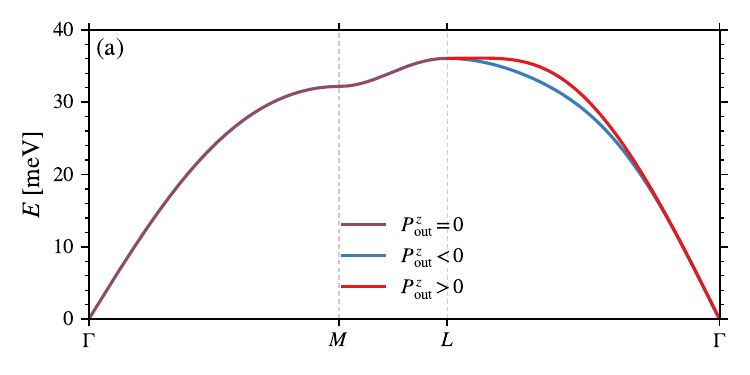} 
    \includegraphics[width=\columnwidth]{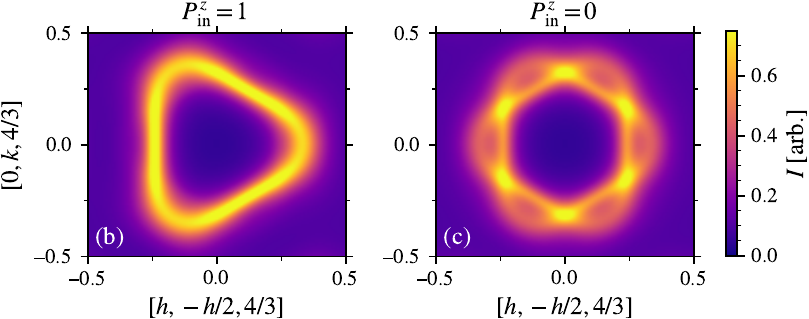} 
    \caption{(a) Magnon dispersion of MnTe using the parameters from \citet{liu2024} along a path $\Gamma$-$M$-$L$-$\Gamma$ in momentum space. The chirality of the magnon bands, which determines the polarization of outgoing neutrons $P^z_{\rm out}$, is indicated by the color of the bands and is visible along the $\Gamma$-$L$ line, cancelling due to symmetry along the other segments of the path. (b,c) Comparison of neutron scattering intensity in the $k_x$-$k_y$ plane with $k_z = 4/3(2\pi/c)$ at $E = 33\ {\rm meV}$ for polarized ($P^z_{\rm in}=1$) and unpolarized ($P^z_{\rm in}=0$) incoming neutrons. Intensity is averaged with a (Lorentzian) energy broadening $\delta E = 1\ {\rm meV}$ to loosely mimic finite experimental resolution.
    }
    \label{fig:mnte}
\end{figure}

We now take an explicit example: that of insulating MnF$_2$. As Mn$^{2+}$ has quenched orbital angular momentum $L=0$ and spin $S=5/2$ we expect the principal magnetic couplings to be highly isotropic. A minimal model for the altermagnetism in this material could include nearest neighbor exchange with a coupling $\Jo$ between the two magnetic sublattices. With this coupling alone, the modes are doubly degenerate, with $\epsilon_{\vec{k},1} = \epsilon_{\vec{k},2} = 8\Jo S(1-\gamma_{\vec{k}}^2)^{1/2}$ and
$C_{\vec{k}} = [(1-\gamma_{\vec{k}})/(1+\gamma_{\vec{k}})]^{1/2}$ where $\gamma_{\vec{k}}= \cos (ak_x/2)\cos (ak_y/2)\cos (ck_z/2)$. This (accidental) degeneracy originates from the fact that the nearest neighbor coupling has a higher symmetry than the lattice~\cite{Gohlke2023,dagnino2024}. The shortest range exchange that has the true symmetry of the lattice appears with representative bond direction along $a(\vhat{x}+\vhat{y})$. At this distance there are two inequivalent couplings $\Jt$ and $\Jtp$ which, when different, induce a finite $\delta {\F}_{\vec{k}}$ lifting the degeneracy of the two magnon bands. Explicitly one finds that
\begin{subalign}
\label{swt:mnf2}
{\F}_{\vec{k}}  &= 8S\Jo + 2S(\Jt+\Jtp)[\cos(ak_x)\cos(ak_y) - 1 ], \\
\delta {\F}_{\vec{k}}  &= 4S(\Jt-\Jtp)\sin (ak_x) \sin(ak_y), \\
{\G}_{\vec{k}} &= -8 S \Jo \gamma_{\vec{k}}. 
\end{subalign}
The degeneracy of the modes is therefore lifted over much of the zone except in the $[h0l]$ and $[0kl]$ planes and at the zone boundaries where $\delta {\F}_{\vec{k}}$ vanishes. Note that $\delta {\F}_{\vec{k}}$ changes sign between the quadrants $k_x k_y > 0 $ and $k_x k_y < 0$ reversing the chirality of the modes at a given energy. This is the magnonic signature of a $d$-wave altermagnet. 

A similar exercise can be carried out for the MnTe magnetic structure which is a $g$-wave altermagnet. As before, we couple the different magnetic sublattices via Heisenberg exchange to nearest neighbor. Then we include the shortest range couplings that break the symmetry down to that of the lattice. Ref.~\cite{dagnino2024} contains a systematic study of the couplings that generate altermagnetism from which we read off couplings $\Jf$ exchanges along the representative bond $\vec{r}_1 \equiv (1,2,1)$ with three-fold symmetry and inversion giving six bonds $\pm \vec{r}_1$, $\pm \vec{r}_2$, $\pm \vec{r}_3$ with the same coupling. There is a second symmetry-inequivalent coupling $\Jfp$ with the same bond length as the $\Jf$ bond but with representative $\vec{r}'_1 \equiv (1,2,-1)$ and six equivalent bonds $\pm \vec{r}'_1$, $\pm \vec{r}'_2$, $\pm \vec{r}'_3$ under symmetry. Defining 
$\gamma_{\vec{k}} \equiv \sum_{\mu=1}^3 \cos(\vec{k}\cdot\vec{r}_\mu)$ and 
$\gamma'_{\vec{k}} \equiv \sum_{\mu=1}^3  \cos(\vec{k}\cdot\vec{r}'_\mu)$
one finds
 \begin{subalign}
 \label{swt:mnte}
{\F}_{\vec{k}}  &= 2S\Jo + S(\Jf+\Jfp)(\gamma_{\vec{k}}+\gamma'_{\vec{k}}-6), \\
\delta {\F}_{\vec{k}}  &= 2S(\Jf-\Jfp)(\gamma_{\vec{k}}-\gamma'_{\vec{k}}), \\
{\G}_{\vec{k}} &= 2S\Jo\cos(ck_z/2).
\end{subalign}
as the role of $J_{121}$ and $J_{121}'$ swap in going from sublattice ${\sf A}$ to sublattice ${\sf B}$. 
This has degeneracies at $[hk0]$, the planes $[h0l]$ and at the zone boundaries where $\delta \F_{\vec{k}}=0$. The chiralities reverse in the upper (lower) energy band under a six-fold rotation so the model exhibits $g$-wave altermagnetism in the magnon bands. The model of \citet{liu2024} includes other short-range exchanges (e.g. $J_2$, $J_3$); while we have neglected these in Eq.~(\ref{swt:mnte}) for simplicity they are included in the explicit calculations shown in Fig.~\ref{fig:mnte}.

While we have focused on the case of two magnetic sublattices where the magnons are theoretically simple, the chirality remains well-defined whenever there is a global $U(1)$ symmetry. This thus still leads to positive and negative chirality blocks in the linear spin wave Hamiltonian. Altermagnetism in such bipartite crystal structures with more than two sublattices thus should be detectable using polarized neutrons in a similar manner.

{\it Detecting polarized altermagnetic magnons} $-$
We have seen that while unpolarized neutrons 
can detect the altermagnetic splitting in the 
energy bands, they are not sensitive to the polarization of the modes. 
However, as was the case for domains in the elastic scattering, the chirality of these altermagnetic magnons \emph{can} be detected using polarized neutrons. We will describe two complementary experimental setups to observe this polarization. 

First one can consider the total scattering intensity of an initial polarized beam with initial polarization $\Pin$~\cite{moon1969,lovesey}
$$
\left(\frac{d^2\sigma}{d\Omega d\omega}\right) \propto
\int dt e^{-i\omega t} \left[
\avg{\vec{M}^{\perp}_{-\vec{k}} \cdot \vec{M}^{\perp}_{\vec{k}}(t)}
+i\Pin \cdot
\avg{\vec{M}^{\perp}_{-\vec{k}} \times \vec{M}^{\perp}_{\vec{k}}(t)}
\right],
$$
where we have defined 
$
\vec{M}^{\perp}_{\vec{k}} \equiv 
\vhat{k} \times ( \vec{M}_{\vec{k}} \times \vhat{k})
$
with $\vec{M}_{\vec{k}} \equiv \sum_{\vec{r},n} \mu_n f_n(\vec{k}) e^{i\vec{k}\cdot (\vec{r}+\vec{\delta}_{n})} \vec{S}_{\vec{r},n}$ being  the magnetization operator at wave-vector $\vec{k}$ (assuming isotropic $g$-factors). 
The polarization dependent part couples to an anti-symmetric combination of the moment operators and thus allows access to the anti-symmetric components of the dynamical structure factor.~\footnote{Note that this second term vanishes if $\Pin \cdot \vhat{k} = 0$, as is commonly the case for experiments measuring spin-flip (SF) and non-spin-flip (NSF) scattering, since $\vec{M}^{\perp}_{\vec{k}}$ is orthogonal to $\vhat{k}$ and $\vec{M}^{\perp}_{-\vec{k}} \times \vec{M}^{\perp}_{\vec{k}} \propto \vhat{k}$. The SF and NSF parts in this configuration thus only contain information about the symmetric part of the dynamical structure factor.}

Alternatively, an initially unpolarized beam will become polarized by scattering from the altermagnetic magnons with the outgoing polarization $\Pout$ given by
$$
\Pout = 
\left(\frac{d^2\sigma}{d\Omega d\omega}\right)^{-1}_{\Pin = 0}
\int dt e^{-i\omega t} 
\left[-i\avg{\vec{M}^{\perp}_{-\vec{k}} \times \vec{M}^{\perp}_{\vec{k}}(t)} \right].
$$
This probes the \emph{same} anti-symmetric part of the dynamical structure that is accessible using the initially polarized beam.

Both of these quantities can be readily computed via spin-wave theory. The total intensity due to a polarized beam at zero temperature can be written in terms of weights associated with each magnon band as
$$
\left(\frac{d^2\sigma}{d\Omega d\omega}\right) \propto \sum_{n} \mathcal{W}_{\vec{k},n}(\Pin) \delta(\omega-\epsilon_{\vec{k},n}).
$$
The weights $\mathcal{W}_{\vec{k},n}(\Pin)$ are defined as
$$
\mathcal{W}_{\vec{k},n}(\Pin) = |\vec{\Ap}^{\perp}_{\vec{k},n}|^2
+ i \Pin \cdot (
\vec{\Ap}^{\perp}_{\vec{k},n} \times
[\vec{\Ap}^{\perp}_{\vec{k},n}]^{*}),
$$
in terms of 
$
\vec{\Ap}^{\perp}_{\vec{k},n} = \vhat{k} \times (
\vec{\Ap}_{\vec{k},n} \times \vhat{k}),
$
and where 
$$
\Ap^{\mu}_{\vec{k},n} 
\equiv \sum_{m} 
\left(
\hat{e}^\mu_{m,-} V^m_{\vec{k},n,+}+
\hat{e}^\mu_{m,+}V^m_{\vec{k},n,-}
\right),
$$
where $m$ here runs over the magnetic atoms.
Here we have split our Bogoliubov eigenvectors~\cite{supp} in blocks as $\trp{\vec{V}}_{\vec{k},n} \equiv (\trp{\vec{V}}_{\vec{k},n,+},\trp{\vec{V}}_{\vec{k},n,-})$ and denoted choice of Cartesian frame $\vhat{e}_{n,\pm} \equiv (\vhat{x}_n \pm i\vhat{y}_n)/\sqrt{2}$ for the $n^{\rm th}$ magnetic ion. For the simple two sublattice case discussed above, these vectors would be $\trp{\vec{V}}_{\vec{k},1}=(u_{\vec{k}},0,0,{v}_{\vec{k}})$ and $\trp{\vec{V}}_{\vec{k},2}=(0,u_{\vec{k}},v_{\vec{k}},0)$.
Similarly we can express the polarization of the outgoing intensity at each of the magnon bands due to an initially unpolarized beam as
$$
\Pout(\vec{k},\epsilon_{\vec{k},n})
= -i\frac{\vec{\mathcal{A}}^{\perp}_{\vec{k},n} \times\cc{[\vec{\Ap}^{\perp}_{\vec{k},n}]}}{
\vec{\Ap}^{\perp}_{\vec{k},n} \cdot [\vec{\Ap}^{\perp}_{\vec{k},n}]^{*}
}.
$$

For the case of a colinear altermagnet with two sublattices and isotropic interactions, one can show that
\begin{align*}
    -i\vec{\mathcal{A}}^{\perp}_{\vec{k},n} \times[\vec{\Ap}^{\perp}_{\vec{k},n}]^{*} &= -(-1)^{n}  \vhat{k} (\vhat{k}\cdot \vhat{N}) C_{\vec{k}}, \\
    \vec{\mathcal{A}}^{\perp}_{\vec{k},n} \cdot [\vec{\Ap}^{\perp}_{\vec{k},n}]^{*}
    &= \frac{1}{2}\left[1+(\vhat{k}\cdot\vhat{N})^2\right] C_{\vec{k}},
\end{align*}
where $C_{\vec{k}}$ is the one magnon intensity defined in Eq.~(\ref{eq:defn-c}). From these expressions we see that for an initially polarized beam
the outgoing intensity is modulated by the polarization oppositely for the two bands
$$
\mathcal{W}_{\vec{k},n}(\Pin) = 
\left(
\frac{1}{2}[1+(\vhat{k}\cdot\vhat{N})^2]
+ (-1)^{n} (\Pin \cdot\vhat{k})
(\vhat{k}\cdot \vhat{N})
\right) C_{\vec{k}}.
$$
 One may carry out such an experiment on an instrument in which the incident (or scattered) neutron beam is polarized using the method often referred to as the ``half-polarized'' or the ``flipping ratio method''. One gains the most insight by aligning the neutron polarization along the moment direction with $\Pin = \pm \vhat{N}$. One measures the cross section for $\Pin = +\vhat{N}$ and again for $-\vhat{N}$ and this gives access to the relevant terms. No polarization analysis is needed in this case; explicitly, if we define
$\Delta \mathcal{W}_{\vec{k},n} \equiv
[\mathcal{W}_{\vec{k},n}(+\vhat{N})-
\mathcal{W}_{\vec{k},n}(-\vhat{N})]/2$ and divide out the unpolarized result one directly arrives at
$$
\frac{\Delta \mathcal{W}_{\vec{k},n}}{\mathcal{W}_{\vec{k},n}(\vec{0})} = (-1)^{n}\left(
\frac{2
(\vhat{k}\cdot \vhat{N})^2}
{1+(\vhat{k}\cdot\vhat{N})^2}\right),
$$
independent of the one magnon intensity factor $C_{\vec{k}}$.

The differing signs for the two modes directly correspond to their respective chiralities. Note that in situations where there is altermagnetic splitting for scattering wavevectors along the moment direction (which is certainly the case for ideal altermagnets), the entire scattering intensity for such wavevectors can be reweighted from one mode to the other simply by reversing the polarization direction of the in-going beam. 

Similarly for an initially unpolarized beam the
polarization of the intensity of the magnon bands is opposite on each band
$$
\Pout(\vec{k},\epsilon_{\vec{k},n})
= -(-1)^n \left(\frac{2(\vhat{k}\cdot \vhat{N})}{
1+(\vhat{k}\cdot \vhat{N})^2}\right)\vhat{k}.
$$
For the component of the polarization along $\vhat{N}$ this yields the same result (up to sign) as ${\Delta \mathcal{W}_{\vec{k},n}}/{\mathcal{W}_{\vec{k},n}}(\vec{0})$. In this case, the experiments require use of a neutron polarization analyser (for example CRYOPAD \cite{cryopad1,cryopad2}). 

Examples of what experimental neutron scattering intensities might look like are shown in Figs.~\ref{fig:mnf2} and \ref{fig:mnte}. For MnF\textsubscript{2} (Fig.~\ref{fig:mnf2}) we show the outgoing polarization of the intensity for an initially unpolarized neutron beam. The polarization directly shows the $d$-wave symmetry of the magnon chirality in both the dispersion and fixed energy cuts. For MnTe (Fig.~\ref{fig:mnte}) we show the dispersion (with magnon chiralities indicated) and the neutron intensity for an experiment where the incoming neutrons are polarized. Compared to the unpolarized intensity, the polarization enhances the intensity of one chiral band, extinguishing the other and thus revealing the spatial $g$-wave structure of the magnon wave-functions. For a smaller altermagnetic splitting (or broader energy resolution) this extinguishment will not be perfect, but a spatial anisotropy in the intensity should still be visible.

{\it Weakly breaking the $U(1)$ symmetry} $-$ So far we have considered only ideal altermagnets where there is a residual $U(1)$ symmetry in the magnetically ordered state resulting in magnon modes with well-defined chirality. In altermagnetic materials, this symmetry is expected to hold to an excellent approximation. Weak anisotropic couplings will inevitably that break this $U(1)$ down to the crystalline symmetries via (e.g.) the magnetostatic dipolar interaction or spin-orbit coupling. How does this affect chirality as a signature of altermagnetic magnons?

Start in the limit where the magnons are chiral and degenerate -- for example in a conventional Heisenberg antiferromagnet or in an altermagnet at wavevectors where the degeneracy is protected by spin-space symmetries -- and thus $\delta {\F}_{\vec{k}}=0$. Anisotropies breaking the $U(1)$ symmetry modify the spin wave Hamiltonian through the addition of a $\delta \mat{M}_{\vec{k}}$ of the form
$$
\delta \mat{M}_{\vec{k}} = 
  \left(
    \begin{array}{cccc}
    0& {\F}^{\sA\sB}_{\vec{k}} & {\G}^{AA}_{\vec{k}} & 0 \\
    \cc{[\F^{\sA\sB}_{\vec{k}}]} & 0& 0 & {\G}^{\sB\sB}_{\vec{k}} \\
    \cc{[\G^{\sA\sA}_{\vec{k}}]} & 0 &0 & \cc{[\F^{\sA\sB}_{\vec{k}}]} \\
    0 & \cc{[\G^{\sB\sB}_{\vec{k}}]} & {\F}^{\sA\sB}_{\vec{k}} & 0
    \end{array}
    \right).
$$
We assume that the perturbation is non-altermagnetic in the sense that it does not distinguish between the two sublattices and thus does not contribute to $\delta \F_{\vec{k}}$.~\footnote{We note that anisotropic interactions generally realize the symmetry of the lattice even at short-range. If the dominant short-range interactions are from sublattice $\sf A$ to $\sf B$ then this can at best only generate a constant contribution to $\delta \F_{\vec{k}}$; wave-vector dependence must originate from intra-sublattice couplings.}
Any (perturbative) changes to ${\F}_{\vec{k}}$, ${\G}_{\vec{k}}$ will be absorbed into their definitions. 

Projecting $\delta \mat{M}_{\vec{k}}$ into the
subspace spanned by the eigenvectors $\vec{V}_{\vec{k},1}$ and $\vec{V}_{\vec{k},2}$ we obtain the effective Hamiltonian
$$
\left(\begin{array}{cc}
    0 & 
    t_{\rm eff}  \\
     \cc{t}_{\rm eff} & 0
\end{array}
\right),
$$
where we have defined
$t_{\rm eff} \equiv 
\h{\vec{V}}_{\vec{k},1} \delta \mat{M}_{\vec{k}} \vec{V}_{\vec{k},2}$. The corrected eigenvectors are linear combinations of $\vec{V}_{\vec{k},1}$ and $\vec{V}_{\vec{k},2}$ of the form
$$
\frac{1}{\sqrt{2}}\left(
\vec{V}_{\vec{k},1} \pm e^{-i\phi}
\vec{V}_{\vec{k},2}
\right),
$$
where $t_{\rm eff} \equiv |t_{\rm eff}|e^{i\phi}$. Correspondingly we have the new $\vec{\mathcal{A}}_{\vec{k},\pm}$ and the the polarization dependent contribution is then proportional to
$$
i\vec{\Ap}^{\perp}_{\vec{k},\pm} \times [\vec{\Ap}^{\perp}_{\vec{k},\pm}]^{*} = 
\pm \frac{1}{2}\left(i e^{-i\phi} \vec{\Ap}^{\perp}_{\vec{k},2} \times
[\vec{\Ap}^{\perp}_{\vec{k},1}]^{*}
-{\rm c.c.}\right),
$$
since $\vec{\mathcal{A}}_{\vec{k},1} \times [\vec{\Ap}^{\perp}_{\vec{k},1}]^{*} = -\vec{\mathcal{A}}_{\vec{k},2} \times [\vec{\Ap}^{\perp}_{\vec{k},2}]^{*}$. Then since we have $\vec{\mathcal{A}}_{\vec{k},1} \propto [\vec{\Ap}^{\perp}_{\vec{k},2}]^{*}$ their cross product vanishes. Thus small $U(1)$ breaking anisotropies, absent a finite $\delta {\F}_{\vec{k}}$, tend to maximally mix the chiral modes leading to a vanishing polarization dependent part of the neutron scattering cross section.

When both $U(1)$ breaking terms and $\delta {\F}_{\vec{k}}$ are finite, but small the perturbative analysis follows standard first order degenerate perturbation theory leading to an avoided crossing with splittings $\delta \epsilon_{\vec{k},\pm} = \pm 2[h_{\rm eff}^2+|t_{\rm eff}|^2]^{1/2}$ where $h_{\rm eff} \propto \delta A_{\vec{k}}$. The polarization of each mode is then given by $\pm h_{\rm eff}/[h_{\rm eff}^2+|t_{\rm eff}|^2]^{1/2}$, vanishing when $h_{\rm eff}=0$ and equal to $\pm 1$ when $t_{\rm eff}=0$. For small $t_{\rm eff}$ we have polarizations $\pm (1 - |t_{\rm eff}|^2/h_{\rm eff} + \cdots)$ and thus the neutron polarization signatures discussed in this work are perturbatively stable.

{\it Conclusions} $-$ We have shown that polarized neutron scattering has attractive features for the detection and characterization of altermagnets. In particular, elastic scattering with polarization analysis can be used to infer the domain composition of altermagnets. In samples that have been established to have a majority domain, we have described how polarized inelastic neutron scattering can provide a direct, unambiguous experimental signature of the altermagnetic chirality splitting of magnons. This could be used, for example, to reveal the $g$-wave altermagnetism in MnTe. While we have focused on local moment models of altermagnetism, we expect identical polarization signatures from magnon scattering in itinerant altermagnets. Potential future applications of such tools could be in probing features arising from magnon-magnon or magnon-electron interactions in altermagnets~\cite{ReyesOsorio2023,Garcia2024,Costa2024} through their polarization response. Insights from polarized neutrons can thus pave the way for the clear confirmation of altermagnetism in candidate materials and promise to become a central tool in this emerging field. 

\begin{acknowledgments}
PM acknowledges financial support from the CNRS and useful discussions with Dalila Bounoua, Philippe Bourges, Fran\c{c}oise Damay, Quentin Faure and Sylvain Petit. Work at the University of Windsor (JGR) was funded by the Natural Sciences and Engineering Research Council of Canada (NSERC) (Funding Reference No. RGPIN-2020-04970). 
\end{acknowledgments}

\bibliography{references}


\appendix

\section{Review of polarized neutron scattering cross sections}

This section contains a short self-contained review of neutron cross sections from magnetic crystals when either the beam is polarized or the detectors are sensitive to polarization. See also Refs.~\cite{moon1969,lovesey} for a more comprehensive treatment.

The in-going neutron beam will be assumed to have polarization $\Pin \equiv 2 \langle \vec{I} \rangle$ where $\vec{I}$ is the neutron spin. Therefore $0\leq \vert\Pin \vert\leq 1$ where $\vert\Pin \vert=0$ describes an unpolarized beam and $\vert\Pin \vert=1$ is a fully polarized beam along $\hatPin$. The polarization of scattered neutrons is denoted as $\Pout$. 

The ions in the crystal are indexed by $\vec{r}_{n} = \vec{r}+\vec{\delta}_n$ where $\vec{r}$ are the primitive lattice vectors of the magnetic unit cell and $\vec{\delta}_n$ runs over ions in the this unit cell. Each atom carries a moment $\vec{\mu}_{\vec{r},n} = g_n \mu_{\rm B} \vec{S}_{\vec{r},n}$ (that are zero for the non-magnetic ions). We introduce $b_n$ for the coherent scattering length of the ions. 

For elastic scattering of neutrons the cross section can be broken into three parts
\beq
\frac{d\sigma}{d\Omega} = \left( \frac{d\sigma}{d\Omega} \right)_{\rm nuc} + \left( \frac{d\sigma}{d\Omega} \right)_{\rm mag} + \left( \frac{d\sigma}{d\Omega} \right)_{\rm nuc-mag},
\eeq
the cross section from nuclear scattering of unpolarized neutrons, the part coming from scattering of unpolarized neutrons from magnetic moments and the nuclear-magnetic interference term. The last term can be written
\beq
\left( \frac{d\sigma}{d\Omega} \right)_{\rm nuc-mag} \propto \sum_{\vec{r},\vec{r}'} e^{i\vec{k}\cdot (\vec{r} - \vec{r}')} \Pin^\perp \cdot {\rm Re}\left[ \vec{F}^*_{\rm mag}(\vec{k}) F_{\rm nuc}(\vec{k}) \right]
\label{eq:PolNM}
\eeq
where we have define the form factors
\begin{align}
& F_{\rm nuc}(\vec{k})  \equiv \sum_{n} e^{i\vec{k}\cdot\vec{\delta}_n} b_n \\
& \vec{F}_{\rm mag}(\vec{k})  \equiv  \mu_{\rm B}  \sum_{n} e^{i\vec{k}\cdot\vec{\delta}_n} g_n f_n(\vec{k}) \avg{\vec{S}_{n}} = \avg{\vec{M}_{\vec{k}}},
\end{align}
where the static moment does not depend on the unit cell $\avg{\vec{S}_{\vec{r},n}} \equiv \avg{\vec{S}_n}$. We have adopted the notation $\Pin^\perp  \equiv \hat{\vec{k}} \times (\Pin \times \hat{\vec{k}})$ for any vector quantity that appears below. 

This reduces to Eq.~(\ref{eq:NMmix}) when the moments are colinear. This contribution vanishes when the scattering wavevector is aligned with the in-going polarization direction or when the moments are colinear and aligned with $\vec{k}$. Of particular interest to us is the fact that the contribution of Eq.~(\ref{eq:PolNM}), out of the entire elastic cross section, is time reversal odd so it can potentially distinguish domains related by time reversal symmetry.  

The purely magnetic scattering cross section can also be split into the usual unpolarized part and a polarized part  
$$
\left( \frac{d\sigma}{d\Omega} \right)_{\rm mag}=\left( \frac{d\sigma}{d\Omega} \right)_{\rm mag, unpol}+\left( \frac{d\sigma}{d\Omega} \right)_{\rm mag, pol}.
$$
The purely magnetic polarized contribution takes the form
\beq
\left( \frac{d\sigma}{d\Omega} \right)_{\rm mag, pol} \propto  -i \sum_{\vec{r},\vec{r}'} e^{i\vec{k}\cdot (\vec{r} - \vec{r}')}   
\Pin \cdot  \cc{[\vec{F}^{\perp}_{\rm mag}(\vec{k})]}  \times
\vec{F}^\perp_{\rm mag}(\vec{k}).
\eeq
Since this part is time reversal invariant it does not contain information about the domain populations.

The cross section for inelastic scattering of neutrons can be decomposed in an similar way. Here we consider only the purely magnetic contributions, starting with the usual unpolarized case
\begin{equation}
    \left( \frac{d^2\sigma}{d\Omega d\omega} \right)_{\rm mag, unpol} \propto   \int_{-\infty}^{+\infty} dt e^{-i\omega t} \avg{ \vec{M}^{\perp}_{-\vec{k}}(0)\cdot \vec{M}^{\perp}_{\vec{k}}(t)},
\end{equation}
and the polarization dependent contribution
\beq
\left( \frac{d^2\sigma}{d\Omega d\omega} \right)_{\rm mag, pol} \propto  +i \int_{-\infty}^{+\infty} dt e^{-i\omega t}   \Pin\cdot \avg{\vec{M}^{\perp}_{-\vec{k}}(0) \times \vec{M}^{\perp}_{\vec{k}}(t)},
\eeq
which vanishes when the beam is unpolarized. Due to the cross products appearing in this expression we see that polarized neutrons evidently allow one to measure components of the antisymmetric spin-spin correlator. 

Now we restrict our attention to the cross section for altermagnets in the zero spin-orbit coupling limit. The moments are assumed to be colinear along the $\vhat{z}$ axis and we suppose that the Hamiltonian has a global $U(1)$ symmetry such that the magnons have a well-defined chirality. If we also put the beam polarization along the moment direction the cross section is 
\begin{widetext}
\beq
\frac{d^2\sigma}{d\Omega d\omega} \propto \frac{1}{2}\int_{-\infty}^{+\infty} dt e^{-i\omega t} \left[ 
 \left( \left\langle M^+_{-\vec{k}}(0)M^-_{\vec{k}}(t)  \right\rangle + \left\langle M^-_{-\vec{k}}(0)M^+_{\vec{k}}(t)  \right\rangle \right)
+ P^z_{\rm in}(\hat{k}_z)^2  
 \left( \left\langle M^+_{-\vec{k}}(0)M^-_{\vec{k}}(t)  \right\rangle - \left\langle M^-_{-\vec{k}}(0)M^+_{\vec{k}}(t)  \right\rangle \right)
\right].
\eeq
The dynamical correlators $\left\langle M^+_{-\vec{k}}(0)M^-_{\vec{k}}(t)  \right\rangle$ in this expression is non-vanishing for one chirality and vanishes for the other and {\it vice versa} for the other correlator. 

Having considered the cross section for magnetic scattering of polarized neutrons we now examine expressions for the polarization of the scattered beam. This polarization is written as the ratio of a correlator with intensity for the unpolarized case
\begin{equation*}
\Pout = \left(\frac{d^2\sigma}{d\Omega d\omega}  \right)_{\Pin=0}^{-1} \int_{-\infty}^{+\infty} dt e^{-i\omega t} \left[ \left\langle \vec{M}^\perp_{-\vec{k}}(0) \left( \Pin \cdot \vec{M}^\perp_{\vec{k}}(t) \right) \right\rangle + \left\langle  \left( \Pin \cdot \vec{M}^\perp_{-\vec{k}}(0) \right) \vec{M}^\perp_{\vec{k}}(t) \right\rangle - \Pin \left\langle \vec{M}^\perp_{-\vec{k}}(0) \cdot \vec{M}^\perp_{\vec{k}}(t)  \right\rangle  
- i \left\langle \vec{M}^\perp_{-\vec{k}}(0) \times \vec{M}^\perp_{\vec{k}}(t)  \right\rangle  
\right].
\end{equation*}
Restricting our attention to altermagnetism again, suppose the moments are aligned along $\vhat{z}$ and have a $U(1)$ symmetry and for simplicity that the ingoing beam is unpolarized. Then the outgoing polarization is given by 
\beq
\Pout = +\frac{1}{2} \vhat{k}\hat{k}_z \left(\frac{d^2\sigma}{d\Omega d\omega}  \right)_{\Pin=0}^{-1} \int_{-\infty}^{+\infty} dt e^{-i\omega t}  \left[  
 \left\langle M^-_{-\vec{k}}(0)  M^+_{\vec{k}}(t)  \right\rangle  - \left\langle M^+_{-\vec{k}}(0)  M^-_{\vec{k}}(t)  \right\rangle
\right].
\eeq
We thus see that the outgoing polarization is directly related to the magnon chirality.

\end{widetext}

\section{Review of linear spin-wave theory}
Here we briefly review linear spin-wave theory to establish notation used in the main text. We consider
a spin Hamiltonian
\begin{equation*}
    H = \sum_{\vec{r}\vec{\delta}} \sum_{nn'}\trp{\vec{S}}_{\vec{r},n}  \mat{J}_{\vec{\delta},nn'} \vec{S}_{\vec{r}+\vec{\delta},n'}
\end{equation*}
where $\vec{r},n$ denotes the unit cell and sublattice index of the spin.
We now consider a semi-classical expansion about some magnetically ordered ground state. The spin operators can expressed in terms of Holstein-Primakoff bosons as
\begin{align*}
    \vec{S}_{\vec{r},n} &\equiv
  \sqrt{S}\left[
    \left(1-\frac{\h{a}_{\vec{r},n}a_{\vec{r},n}}{2S}\right)^{1/2} \nh{a}_{\vec{r},n}  \vhat{e}_{n,-}+
    \h{a}_{\vec{r},n}  \left(1-\frac{\h{a}_{\vec{r},n}a_{\vec{r},n}}{2S}\right)^{1/2}
    \vhat{e}_{n,+}\right]\\
  &+\left(S-\h{a}_{\vec{r},n}a_{\vec{r},n}\right)\vhat{e}_{n,0},
\end{align*}
where the vectors
$\vhat{e}_{n,\pm}$, $\vhat{e}_{n,0}$ define a local frame of reference; in a more conventional
Cartesian basis one defines $\vhat{e}_{n,\pm} \equiv (\vhat{x}_n \pm i\vhat{y}_n)/\sqrt{2}$ and
$\vhat{e}_{n,0} \equiv \vhat{z}_n$. 
It is useful to write the exchange matrix in the frame aligned with these axes 
\begin{equation}
  \label{eq:lswt:exchange}
\mathcal{J}^{\mu\mu'}_{\vec{\delta},nn'} \equiv \trp{\vhat{e}}_{n,\mu} \mat{J}_{\vec{\delta},nn'}\nh{\vhat{e}}_{n',\mu'},
\end{equation}
with $\mu = 0,\pm$.
Expanding in powers of $1/S$ then yields a semi-classical
expansion about the ordered state defined by $\vhat{e}_{n,0}$.

At order $O(S)$ in the Holstein-Primakoff operators we have
\begin{equation}
  \vec{S}_{\vec{r},n} \approx
  \sqrt{S}\left[
   \nh{a}_{\vec{r},n}  \vhat{e}_{n,-}+
    \h{a}_{\vec{r},n} 
    \vhat{e}_{n,+}\right]
  +\left(S-\h{a}_{\vec{r},n} \nh{a}_{\vec{r},n} \right)\vhat{e}_{n,0},
\end{equation}
Inserting this into our spin Hamiltonian and keeping only terms
to $O(S)$ yields
\begin{equation}
  \label{eq:lswt:ham}
  H = E_0 +\frac{1}{2}  \sum_{\vec{k}}
  \left(\trp{[\h{\vec{a}}_{\vec{k}}]}\ \trp{\vec{a}}_{-\vec{k}}\right)
  \left(
    \begin{array}{cc}
      \mat{A}_{\vec{k}} & \mat{B}_{\vec{k}} \\
      \cc{\mat{B}}_{-\vec{k}} & \cc{\mat{A}}_{-\vec{k}}
    \end{array}
  \right)
  \left(
  \begin{array}{c}
    \nh{\vec{a}}_{\vec{k}} \\
    \h{\vec{a}}_{-\vec{k}}
  \end{array}\right),
\end{equation}
where $E_0$ is the classical energy, $N_s$ is the number of sublattices, $N$ is the total number of
sites and we have defined the Fourier transforms of the bosons as
$a_{\vec{k}n} \equiv N_c^{-1/2} \sum_{\vec{r}}
e^{-i\vec{k}\cdot(\vec{r}+\vec{\delta}_{n})} a_{\vec{r},n}$ where $N = N_c N_s$. 
The matrices $\mat{A}_{\vec{k}}$ and $\mat{B}_{\vec{k}}$ are given by
\begin{subalign}
  A^{nn'}_{\vec{k}} &= S \left(\mathcal{J}^{+-}_{\vec{k},nn'}
  -\delta_{nn'} \sum_{n''} \mathcal{J}^{00}_{\vec{0},nn''}\right), \\
  B^{nn'}_{\vec{k}} &= S\mathcal{J}^{++}_{\vec{k},nn'},
\end{subalign}
where we have defined the Fourier transforms of the local
exchange matrices as
\begin{equation}
  \label{eq:lswt:local}
\mathcal{J}^{\mu\mu'}_{\vec{k},nn'} \equiv \sum_{\vec{d}} \mathcal{J}^{\mu\mu'}_{\vec{d},nn'}e^{i\vec{k}\cdot(\vec{d}+\vec{\delta}_{n'}-\vec{\delta}_{n})} .
\end{equation}

The linear spin-wave Hamiltonian [Eq.~(\ref{eq:lswt:ham})] can be diagonalized by a Bogoliubov transformation. To do
this one diagonalizes the modified matrix~\cite{blaizot:1986}
\begin{align}
  \label{eq:lswt:bogo}
    \left(
    \begin{array}{cc}
      \mat{A}_{\vec{k}} & \mat{B}_{\vec{k}} \\
      -\cc{\mat{B}}_{-\vec{k}} & -\cc{\mat{A}}_{-\vec{k}}
    \end{array}
  \right) \equiv   \mat{\sigma}_3 \mat{M}_{\vec{k}} 
\end{align}
where $\mat{\sigma}_\mu$ denotes the set of block Pauli matrices.
This yields pairs of eigenvectors $\vec{V}_{\vec{k},n}$ and
$\vec{W}_{-\vec{k},n} = \mat{\sigma}_1
\cc{\vec{V}}_{-\vec{k},n}$ with eigenvalues
$+\epsilon_{\vec{k},n}$ and $-\epsilon_{-\vec{k},n}$. These
vectors can be normalized such that~\cite{blaizot:1986}
\begin{align*}
  \label{eq:lswt:norm}
  \h{\vec{V}}_{\vec{k},n} \mat{\sigma}_3 \nh{\vec{V}}_{\vec{k},n'}
  &= +\delta_{nn'},
  &
    \h{\vec{W}}_{-\vec{k},n} \mat{\sigma}_3 \nh{\vec{W}}_{-\vec{k},n'}
  &= -\delta_{nn'},
  \\
    \h{\vec{W}}_{-\vec{k},n} \mat{\sigma}_3 \nh{\vec{V}}_{\vec{k},n'}
  &= 0.
\end{align*}
One can then write the Hamiltonian in terms of diagonalized bosons, $\gamma_{\vec{k},n}$, as~\cite{blaizot:1986}
\begin{align}
  H =
  \sum_{\vec{k},n}\epsilon_{\vec{k},n} \h{\gamma}_{\vec{k},n} \nh{\gamma}_{\vec{k},n}  + {\rm const.}
\end{align}

\end{document}